\documentclass[nofootinbib,showpacs,preprintnumbers,amsmath,amssymb]{revtex4}
\usepackage{epsfig}

\begin{document}

\title{Generalization of the DGLAP for the structure function $g_1$
to the region of small~$x$}

\vspace*{0.3 cm}

\author{B.I.~Ermolaev}
\affiliation{Ioffe Physico-Technical Institute, 194021
 St.Petersburg, Russia}

\author{M.~Greco}
\affiliation{Dept of Physics and INFN, University Rome III, Rome,
Italy}

\author{S.I.~Troyan}
\affiliation{St.Petersburg Institute of Nuclear Physics, 188300
Gatchina, Russia}

\begin{abstract} The explicit expressions for the non-singlet and
singlet components of the DIS structure function $g_1$ comprising
the DGLAP -expressions for the coefficient functions and the
anomalous dimensions, and accounting for the total resummation of
the most singular contributions to those are obtained.
\end{abstract}

\maketitle

\section{Introduction}

The standard theoretical instrument for studying the structure
function $g_1$ is the DGLAP\cite{dglap}. In this approach,
 $g_1(x, Q^2)$ can be
represented as a convolution of the coefficient functions and the
evolved quark distributions calculated with NLO approximation.
Those results, having been completed with appropriate fits for the
initial quark distributions, provide a good agreement with the
available experimental data.

On the other hand, the DGLAP evolution was originally obtained for
operating in the range of rather large $x$ where higher-loop
contributions to the coefficient functions and the anomalous
dimensions are small. It is clear that when $x$ is decreasing,
such corrections are becoming more and more essential and the
DGLAP should stop to work well at certain values of small $x$.

In the present paper we use the results of our previous papers
\cite{egt1,egt2,egt3} to demonstrate that the impact of the
high-order corrections on the $Q^2$ and $x$ -evolutions of the
non-singlet structure functions is quite sizable for $x \leq
10^{-2}$.

\section{Difference between DGLAP and our approach}

As the DGLAP -expressions for the non-singlet structure functions
are well-known.
In order to make the all-order resummation of the
double-logarithmic contributions to $g_1$,
in Refs.~\cite{ber} was used an alternative approach, composing
and solving the Infrared Evolution Equations. This approach was
improved in Refs.~\cite{egt1} where the single-logarithmic
contributions were accounted for and the QCD coupling
was
running while in Refs.~\cite{ber} it was considered as fixed. In
contrast to the DGLAP parametrization $\alpha_s =
\alpha_s(k^2_{\perp})$, Refs.~\cite{egt1} used the other
parametrization.
The argumentation in favor of such a parametrization was
given in   Ref.~\cite{egt2}. In particular, it was shown there
that this new parametrization coincides with the DGLAP
-parametrization when $x$ is not far from 1 but those
parameterizations differ a lot when $x \ll 1$.
Refs.~\cite{egt1} suggest the following
formula for the non-singlet component of $g_1 (\equiv g_1^{NS})$:
\begin{equation}
\label{gnsint} g_1^{NS}(x, Q^2) = (e^2_q/2) \int_{-\imath
\infty}^{\imath \infty} \frac{d \omega}{2\pi\imath }(1/x)^{\omega}
C_{NS}(\omega)
\delta q(\omega) \exp\big( H_{NS}(\omega) y\big)~,
\end{equation}
with $y = \ln(Q^2/ \mu^2)$ so that $\mu^2$ is the starting point
of the $Q^2$ -evolution. The new coefficient function $C_{NS}$ is
expressed in terms of new anomalous dimension $H_{NS}$:
\begin{equation}
\label{cns} C_{NS} =\omega/(\omega - H_{NS}(\omega))
\end{equation}
and the new anomalous dimension  $H_{NS}(\omega)$ accounting for
the total resummation of the double- and single- logarithmic
contributions is

\begin{equation}
\label{hns} H_{NS} = (1/2) \Big[\omega - \sqrt{\omega^2 -
B(\omega)} \Big]
\end{equation}
where
Where $B$ and related to it
the Mellin transform of $\alpha_s$, $A(\omega)$
are expressed in terms of $l =
\ln(1/x)$, $b = (33 - 2n_f)/12\pi$ and the color factors
 (see Ref.~\cite{egt1} for details).

\section{Comparison of the $x$  evolution for
$g_1^{NS}$ by DGLAP to the one by eq.~(\ref{gnsint})}

Let us compare our results (\ref{gnsint}) to the expressions for
  $g_{1~NLO}^{NS}$
obtained with the NLO DGLAP. In order to be independent of fits
for $\delta q$, we use the simplest input, the bare quark one. In
other words, we compare pure evolutions for the non-singlets. We
compare them when $x$ is changing while $Q^2$ is fixed. We define
$R_{NLO}$ as follows:

\begin{equation}
\label{rx} R_{NLO}(x) = g_1^{NS}/g_{1~NLO}^{NS}
\end{equation}
when $Q^2$ is fixed and the initial quark distribution corresponds
to the bare quark. The results for $R_{NLO}(x)$ and are presented
in Fig.~1 taken from Ref.~\cite{egt3}. It shows that for $x \geq
0.005$ the NLO DGLAP evolution predicts the values for the
non-singlets similar to the ones predicted by our evolution.
However, for $x \leq 0.005$ the situation is opposite. Therefore,
we gather that $x \approx 10^{-2}$ is the point where the impact
of higher-loop contributions becomes sizable. However, it is known
that DGLAP actually works successfully at lower values of $x$. We
suggest the following explanation to this fact: it is the impact
of the standard fits for the initial quark density used in DGLAP.
Indeed, all such fits contain the terms singular when $x \to 0$
and therefore they are able to mimic the impact of the higher-loop
contributions absent in DGLAP.

\section{Combining DGLAP with our higher-loop contributions}

Eq.~(\ref{gnsint}) accounts for the total resummation of the
double- and single logarithmic contributions to the non-singlet
anomalous dimension and the coefficient function. They are leading
when $x$ is small but the method we have used does not allow to
account for the other contributions, e.g. constants.One can
neglect them when $x$ is small but they become important when $x$
is not far from 1. On the other hand, such contributions are
accounted in DGLAP where the non-singlet coefficient function
$C_{DGLAP}$ and anomalous dimension $\gamma_{DGLAP}$ are known
with the two-loop accuracy:

\begin{equation}
\label{formdglap} C_{DGLAP} = 1 + \frac{\alpha_s(Q^2)}{2\pi}
C^{(1)}, ~~~\gamma_{DGLAP} =
\frac{\alpha_s(Q^2)}{4\pi}\gamma^{(0)} +
\Big(\frac{\alpha_s(Q^2)}{4\pi}\Big)^2 \gamma^{(1)}
\end{equation}

Therefore, we can borrow from DGLAP formulae the contributions
missing in Eq.~(\ref{gnsint}) by adding $C_{DGLAP}$ and
$\gamma_{DGLAP}$ to the coefficient functions and anomalous
dimensions of Eq.~(\ref{gnsint}). However, both $C_{DGLAP}$ and
$\gamma_{DGLAP}$ contain also the terms already accounted for by
Eqs.~(\ref{gnsint}):

\begin{equation}
\label{series} \widetilde{C}_{NS} = 1 + B/(4\omega^2)~,
 ~~~\widetilde{H}_{NS} = B/(2\omega) + B^2/(16\omega^3) ~.
\end{equation}
Now let us make the new coefficient functions  $\hat{C}_{NS}$ and
new anomalous dimensions $\hat{H}_{NS}$ (see Ref.~\cite{egt3} for
more details):

\begin{eqnarray}
\label{newch} \hat{H}_{NS} = \Big[H_{NS} - \widetilde{H}_{NS}\Big]
+ \frac{A(\omega)}{16\pi^2C_F}\gamma^{(0)} +
\Big(\frac{A(\omega)}{16\pi^2C_F}\Big)^2 \gamma^{(1)}, \\
\nonumber \hat{C}_{NS} = \Big[C_{NS} - \widetilde{C}_{NS} \Big] +
 1 + \frac{A(\omega)}{8\pi^2 C_F} C^{(1)}.
\end{eqnarray}

The new coefficient function and anomalous dimension of
Eq.~(\ref{newch}) comprise the total resummation of the leading
contributions from higher loops and the DGLAP expressions in which
$\alpha_s(Q^2)$ is replaced by $A(\omega)$ and therefore can be
used in the regions of small and large $x$. However, the initial
quark densities should be replaced by the new ones, without
singularities when $x \to 0$.

\section{Conclusion}
The total resummation of the most singular logarithmic
contributions to the anomalous dimensions and coefficient
functions leads to the sizable deviation of $g_1$ from the
conventional NLO DGLAP predictions at $x < 10^{-2}$ when the bare
quark input is used for the initial parton density. It also leads
to the Regge asymptotics $\sim x^{-\omega_0}$, with the value of
the intercepts $\omega_0 \approx 0.4$ for the non-singlet $g_1$
and $\omega_0 \approx 0.8$ for the singlet $g_1$. DGLAP
compensates lack of the total resummation by using different kinds
of singular in $x$ fits for the initial parton distributions and
thereby can successfully work at $x < 10^{-2}$. So, our results
can be used for specifying the DGLAP fits, fixing their singular
part.

On the other hand, the DGLAP anomalous dimensions and coefficient
functions contain the terms quite essential at $x$ not far from 1
whereas our approach cannot account for them. In order to get an
approach universally good for large and small $x$, we suggest
combining the NLO DGLAP anomalous dimensions and coefficient
functions and our formulae.

\section{Figure captions}
Fig.~2: $R_{NLO}(x)$ for $g_1^{NS}$ for $Q^2 = 20$~GeV$^2$.

\section{Acknowledgement}
Work supported in part by Grant RSGSS-1124.2003.2.

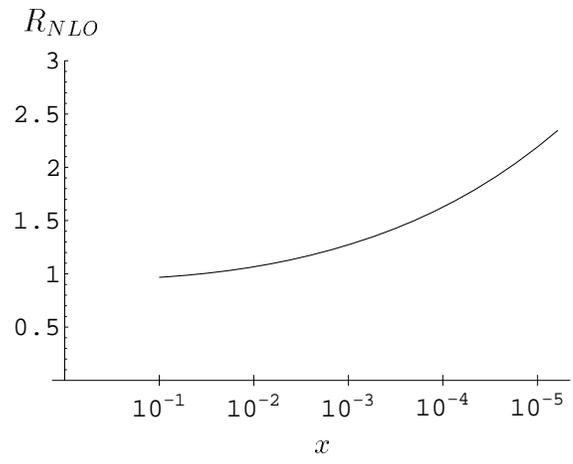
\end{document}